\documentclass[aps,twocolumn,prl,showpacs,superscriptaddress]{revtex4}
\usepackage{graphics}
\usepackage{epsfig}

\begin{document}

\title{{\bf Hierarchy Measures in Complex Networks}}
\author{Ala Trusina}
 \affiliation{Department of Physics, Ume{\aa} University,
90187 Ume{\aa}, Sweden}\affiliation{NORDITA, Blegdamsvej 17, 2100 Copenhagen {\O},
Denmark}
\email{trusina@tp.umu.se}
\author{Sergei Maslov}
\affiliation{ Department of Physics, Brookhaven National Laboratory, \\
Upton, New York 11973, USA}
\email{maslov@bnl.gov}
\author{Petter Minnhagen}
\affiliation{NORDITA, Blegdamsvej 17, 2100 Copenhagen {\O},
Denmark}
\affiliation{Department of Physics, Ume{\aa} University,
90187 Ume{\aa}, Sweden}
\email{minnhagen@nordita.dk}
\author{ Kim Sneppen}
\affiliation{NORDITA, Blegdamsvej 17, 2100 Copenhagen {\O}, Denmark}
\email{sneppen@nbi.dk}

\date{\today}

\begin{abstract}
Using each node's degree as a proxy for its importance, the topological
hierarchy of a complex network is introduced and quantified.
We propose a simple dynamical process used to construct networks
which are either maximally or minimally hierarchical.
Comparison with these extremal cases as well as with random scale-free
networks allows us to better
understand hierarchical versus modular features
in several real-life complex networks.
For random scale-free topologies the extent of topological hierarchy
is shown to smoothly decline
with $\gamma$ -- the exponent of a degree distribution -- reaching its
highest possible value for $\gamma \leq 2$ and quickly approaching zero
for $\gamma>3$.
\end{abstract}

\pacs{89.75.-k, 89.75.Fb}

\maketitle
Networks recently came to the focus of attention of the complex
systems research.
Indeed, most complex systems have an underlying network serving as a
``backbone'' for its dynamical processes.
The large-scale topological organization of a particular
complex network is related to both
its functional role and historical background. Thus it is
important to develop quantitative tools allowing one to detect and
measure significant features in the topology of a given network.\\
The hierarchical organization is a common feature of many
complex systems.
As an example of a hierarchy
one might think of the organizational structure of a large company.
The defining feature
of a hierarchical organization is the existence of a
{\it hierarchical path} connecting any two of its nodes. It can be thought of
as a trajectory of a request initiated at one of the nodes and reaching
its destination node through the ``chain of command''. Such a
request first goes up the steps of the hierarchy until it reaches
the first common boss of its sender and recipient after which it
descend the hierarchical levels down to the destination node.
In most real-life complex systems the simple tree-like hierarchical network
may be augmented by ''shortcuts''
bypassing the chain of command.
Such {\it non-hierarchical shortcuts} make the task of detecting
the hierarchical structure much more non-trivial.
The question we want to address in this work is how to detect and
measure the extent of the hierarchy manifested in the topology of
a given complex network.

Real world networks often have very broad degree distribution
\cite{books}
and thus the degree in itself gives a sensible characteristic
of a node which has to be preserved in any randomized version of a network
\cite{newman2001,maslov2002}.
In fact, in a number of systems nodes with higher degrees
are on average {\it more important} than their lower degree counterparts.
For example, for the Internet
the number of hardwired connections a given Autonomous System
serves as a proxy of its importance with the most connected
nodes being global Internet Service Providers.
For WWW the in-degree of a web page can serve as a measure of its
popularity and hence importance; highly connected
hub-airports of airline networks typically located
in large cities, etc.
In what follows
we use the degree of a node as a proxy
for its rank in the hierarchy based on the
relative importance of nodes.
Thus we propose a way to couple a local topological quantity,
the degree, to a global structure, hierarchy:
Thereby we define and quantify topological hierarchy as
a way to characterize networks beyond their degree distribution
and its two point correlation
function \cite{maslov2002,maslov2002b,newman2}.
However, we would like to point out that our methods could be generalized to
hierarchies defined in terms of any other characteristic of individual nodes,
being it wealth, mass, or some other appropriate quantity.

We quantify the hierarchical topology of a network
using the concept of a {\it hierarchical path} \cite{tang,gao}:
a path between two nodes in a network is called hierarchical
if it consists of
an ``up path" - where one is allowed to step
from node $i$ to node $j$ only if their degrees $k_{i}$, $k_{j}$
satisfy $k_{i}\leq k_{j}$
followed by a ''down path'' - where only steps to nodes of lower
or equal degree are allowed.
Either up or down path is allowed to have zero length.
This definition of a hierarchical path follows the
above mentioned trajectory of a request which is first forwarded up
and then descends down the levels of a hierarchy quantified by $k_i$.
It is also similar to the definition proposed in \cite{tang}
and \cite{gao}.
The length of the shortest
hierarchical path between a given pair of nodes can be either:
1) equal to the length of the shortest path between these nodes;
2) longer than it; 3) not exist at all
if these two nodes cannot be connected by any hierarchical path.
The fraction of pairs in each of these three categories is denoted
as ${\cal F}$, ${\cal S}$, and ${\cal U}=1-{\cal F}-{\cal S}$ correspondingly.
Equivalently, the hierarchical fraction $\cal{F}$ can be viewed as
a fraction of shortest paths
in the network that are hierarchical, while
${\cal S}$ as the probability of
finding a {\it non-hierarchical shortcut} - a path shorter than the shortest hierarchical
path between a pair of nodes.

We analyzed in detail the hierarchical structure of several real-life
complex networks:
1) The Internet consisting of autonomous systems (AS) hardwired to each other \cite{NLANR}
(6474 nodes, 1572 edges);
2) The largest connected component of the yeast protein interaction
network \cite{ito2001} (2839 nodes, 4220 edges);
3) The largest connected component of the E-mail
mutual correspondence network (25151 nodes, 19963 edges)
\cite{email};
4) The network of CEOs (executive company directors); two directors are connected if they
both belong to at least one common board of directors (6193 nodes, 43077 edges) \cite{ceo}.
To have a reference point, we compare these networks with
their random connected counterparts
in which the degree of every individual node is strictly preserved.
In practice this is done by multiple edge-rewiring
moves \cite{maslov2002,maslov2002b} where two links between two randomly selected
pairs of nodes are rewired with the constraint that one only accepts moves in which
no double links are created, and where the network remains connected.
We find that the randomized version of the Internet, characterized by
${\cal F}_r=0.99$, is almost as  hierarchical as the real Internet, where
${\cal F}=0.95$. The same is true for the E-mail network (${\cal
F}_r=0.98$ vs ${\cal F}=0.97$) and the CEO network (${\cal
F}_r=0.84$ vs ${\cal F}=0.78$). On the other hand, the
randomized protein interaction
network, ${\cal F}_r=0.88$, is significantly
more hierarchical than the real one, ${\cal F}=0.33$.
This anti-hierarchical feature of the protein interaction network reflects
a topology where highly connected nodes are placed on the
periphery, and not in the center of the network \cite{maslov2002}.

We also found that a small reduction in ${\cal F}$ for
the Internet compared to its randomized counterpart is
mainly due to an increase in the number of
non-hierarchical shortcuts, ${\cal S}=0.02$ (${\cal S}_r$=0).
This feature is even more pronounced for the yeast protein network,
with ${\cal S}=0.17$ (${\cal S}_r$ = 0.02).
One of the possible explanation for this phenomena can be
the natural tendency toward shorter distances and thus toward faster
and more specific
signaling.

As was shown above, a random network with a given degree distribution provides a
useful benchmark for the extent of hierarchy in real-life complex
networks. It is interesting also to consider the extreme cases:
that is to construct networks that are the most or
alternatively the least hierarchical for a given degree
distribution. This is important for positioning real networks not only with the respect
to its random counterpart, but also with respect to the extreme limits
the network of a given degree distribution can achieve.

Similar to a randomized version, the {\it maximally hierarchal}
version of a network is generated by multiple
rewirings of pairs of edges. One has to add however a
particular preference for reconnection:
At each step one selects two pairs of connected nodes and attempts to
connect the node with the {\it highest} among these four nodes to
the node of the {\it next highest} degree in this subset.
The remaining two nodes are then linked together (multiple
links are forbidden, and the network should always remain
connected).

\begin{figure}[t] 
\centerline{\epsfig{file=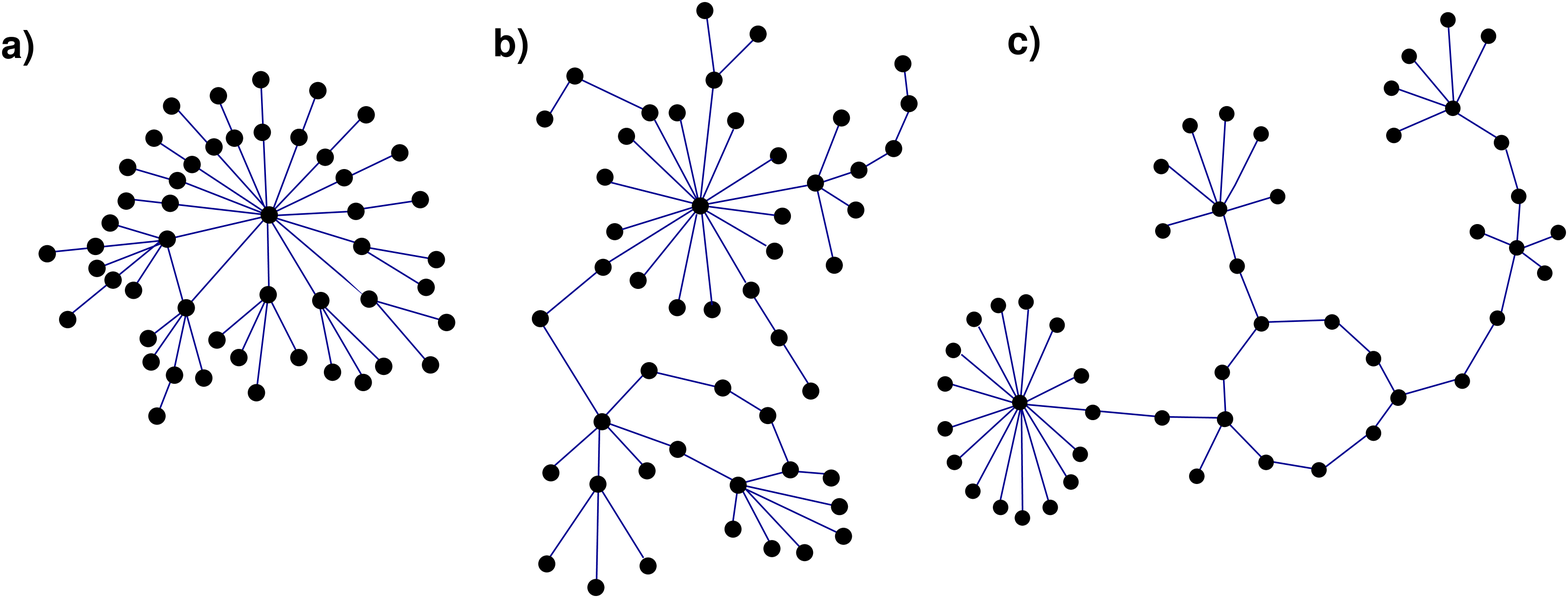,height=0.13\textheight,angle=0}}
\caption{
Maximally hierarchical (a), random (b) and maximal anti-hierarchical (c) networks of size
$N=50$ nodes and node degree distribution
$f(k)\propto 1/k^{2.5}$.}
\label{artWork}
\end{figure}

The {\it maximally anti-hierarchal} version of a network
can be constructed by the same algorithm
but with the opposite preference of reconnection:
the node with the {\it highest} degree is linked
with that with the {\it lowest} degree.
In Figs.~\ref{artWork}a,c we show the maximally hierarchical respectfully anti-hierarchical
networks with the same node degree distribution as a random network, shown in Fig.~\ref{artWork}b.
We have found that applying the above algorithm to all four empirical networks
it is possible to achieve the limits where ${\cal F}$=1 for maximal hierarchy and
${\cal F}$ $\approx 0$ for maximal anti-hierarchy.

As one can see from Fig.~\ref{artWork}, the maximally hierarchical or anti-hierarchical networks
show strong correlations between degrees of connected nodes \cite{redner}
that can be quantified through either the correlation profile \cite{maslov2002}
or the assortativity measure, $r$ introduced in Ref. \cite{newman2}.
We consider a modified assortativity measure,
similar to the one used in Ref.~\cite{newman2}, but here defined as
$r_{AD}=\ln(\frac{\langle k_i k_j\rangle}{\langle \langle k_i k_j\rangle
\rangle_{r}})$, where $\langle k_i k_j\rangle$
is the average over all pairs $i$ and $j$ of nearest-neighbor nodes
in the network and
$\langle \langle k_i k_j
\rangle \rangle_{r}$ is the average of
$\langle k_i k_j\rangle$ in an ensemble of randomized networks
generated as described above \cite{maslov2002,maslov2002b}.
We find the maximally hierarchical topology to
be always assortative ($r_{AD}>0$),
while the maximal anti-hierarchical topology - disassortative ($r_{AD}<0$).
For example, for a network with the node degree distribution
$f(k)\propto 1/k^{2.5}$, $r_{AD}=0.14$ and $-1.24$ for the maximal hierarchy and
the anti-hierarchy  respectively. For comparison,the protein-protein
interactions in yeast, which are well described by $f(k)\propto 1/k^{2.5}$,
has $r_{AD}=-0.82$.
We further stress that assortativity and hierarchical topology are in general
not prerequisites for each other.

Motivated by the abundance of real-life networks characterized
by a broad, often scale-free, degree distribution
(as it is the case for the empirical networks we are considering here),
we quantified the hierarchical fraction ${\cal F}$ as a function of the exponent
$\gamma$ in {\it random} scale-free networks with a power law degree
distribution $f(k)\propto 1/k^{\gamma}$. Such networks
were constructed by first generating a set of power-law
distributed degrees of individual nodes, then linking the
edges to create a single-component network, and finally
randomizing the resulting network using the algorithm of Ref.
\cite{maslov2002b}, which preserves individual degrees and
connectedness of the network.

\begin{figure}[t] 
\centerline{\epsfig{file=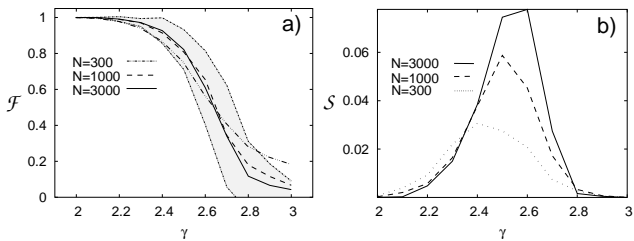,
height=0.13\textheight,angle=0}}
\vskip -0.4cm
\caption{a) The fraction of shortest paths that are also hierarchical,
${\cal F}$ and  b) The fraction of non-hierarchical short-cuts,
${\cal S}$, as a function of $\gamma$ for three system sizes, $N=300,1000,3000$.
The shadowed area in a) corresponds to error estimates for a network of
size $N=3000$.
}
\vskip -0.6cm
\label{F_S_gamma}
\end{figure}

Fig.~\ref{F_S_gamma}a  shows ${\cal F}$ vs $\gamma$ measured in
random scale-free networks for different system sizes.
The decrease from ${\cal F}$$= 1$  to   ${\cal F}$$\sim 0$ happens
as $\gamma$ grows from 2 to 3
with a smooth transition around $\gamma \sim 2.6$ that weakly depends on the system size.
We remark that for $\gamma \leq 2$, $\cal{F}$ $=1$ and is nearly independent on the
upper cutoff of the degree distribution, which is required in this case.

Fig.~\ref{F_S_gamma}b shows ${\cal S}$ vs $\gamma$ in random
scale-free networks. One can see that ${\cal S}\rightarrow 0$ as $\gamma \rightarrow$ 2 and 3.
Indeed, as $\gamma \rightarrow$ 2 the largest hubs become dominant and
the typical distance in a network approaches 2 (almost any pair of nodes are
connected via at least one hub). This makes most shortest paths via
a hub to be hierarchical.
In the limit $\gamma \rightarrow$ 3, the topology of the network is very
close to a tree, which in its turn implies that the number
of alternative paths approaches zero, and thus again ${\cal S}\rightarrow0$.

We have seen that as $\gamma$ approaches $2$ almost all pairs of
nodes tend to have at least one hierarchical paths connecting
them ( ${\cal F }$ $\rightarrow 1$ and hence
${\cal U}==1-{\cal F}-{\cal S}$ $\rightarrow 0$).
The existence of hierarchical
paths connecting most pairs of nodes means that at
the very least the majority of nodes have at least one
neighbor with a degree higher then their own.

Let us first calculate the probability that a given
edge is attached to a node with degree larger than $k$
\begin{eqnarray}
P_{edge}(\geq k)  \propto && \int_k^K k' f(k') dk'  \nonumber \\
 \propto &&
\left \{ \begin{array}{ll}
     1-\left( \frac{k}{K} \right)^{2 - \gamma},  & \mbox {for $ \gamma <2$ }\\
        k^{2-\gamma},     & \mbox {for $ \gamma >2$ }
  \end{array}
\right.
\end{eqnarray}
Here for $\gamma<2$ one can only have a scale-free distributions
below an upper cutoff $K$.
Thus in the absence of degree-degree correlations
the probability that a node of degree $k$ has at least
one neighbor of degree higher than itself is given by
\begin{eqnarray}
P(k_{neighbor}\geq k) & \propto  & (1-P_{edge}(\geq k))^k  \nonumber \\
\propto &&
\left \{ \begin{array}{ll}
     1-\left( \frac{k}{K} \right)^{(2 - \gamma)k},  & \mbox {for $ \gamma <2$ }\\
     1-(1- k^{2-\gamma})^k, & \mbox {for $ \gamma >2$ }
  \end{array}
\right.
\end{eqnarray}
\begin{figure}[t] 
   \centerline{\epsfig{file=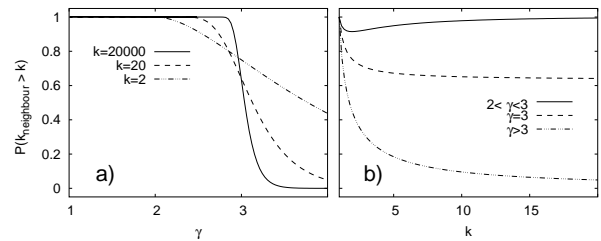,
       height=0.13\textheight,angle=0}}
   \caption{The probability $P(k_{neighbor}\geq k)$ for a node of degree $k$
        to  have a neighbor of degree
        $k_{neighbor}>k$ in an infinitely large scale free network,
 that is not necessarily  connected, as function
 of a) - exponent $\gamma$ and b) - degree of the node $k$ for $\gamma=2.5, 3$ and 4. }
   \label{pNeighbor}
 \end{figure}
In Figure~\ref{pNeighbor} we plot this probability of having a boss
(neighbor with a higher degree) as a function
of $\gamma$ for three different values of the degree $k$, Fig.~\ref{pNeighbor}a,
and as a function of $k$ for different values of $\gamma$, Fig.~\ref{pNeighbor}b.
For $\gamma \leq 2$
both low and high degree nodes
always have a higher connected neighbor
and for $\gamma > 3$ the high $k$ nodes nearly never have a boss.
For $2<\gamma <3$, low connected nodes often have no
higher connected neighbors (see Fig.~\ref{pNeighbor}b), but as
$P(k_{neighbor}>k) \rightarrow 1$ for increasing $k$ there
is a hierarchical core of highly
connected nodes. In popular terms, at these intermediate
values of $\gamma$ many low degree nodes escape the hierarchy, while
medium and highly connected nodes have bosses.
Above $\gamma=3$,
$P(k_{neighbor} > k)$ decreases to zero with degree.
Thus for these high values of $\gamma$ a
network becomes modular with each of the modules
centered around a local hub.

\begin{figure}[t] 
\centerline{\epsfig{file=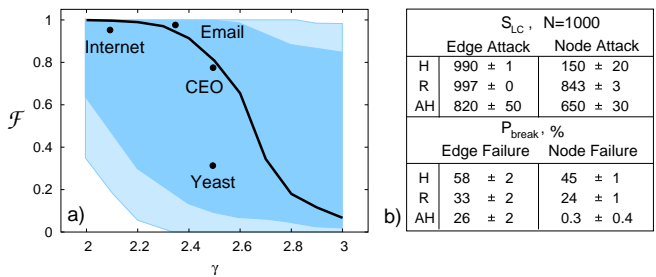,height=0.15\textheight,angle=0}}
\caption{a)
Possible hierarchical organizations of networks, dark-shadowed area shows
the limits for the average values, and light-shadowed area are
the corresponding limits including possible variations from sample
to sample in a random scale-free network with $N=10^3$ nodes.
The black dots show values of the hierarchical fraction ${\cal F}$ and
the degree exponent $\gamma$ for the Internet,
E-mail, Yeast Protein Interaction and CEO networks discussed in the
text. The solid line follows ${\cal F}$ vs $\gamma$ in random
scale-free networks.
b) The results of the robustness analysis for maximally hierarchical (H),
random (R) and anti-hierarchical (AH) scale-free networks with
$\gamma=2.5$.
The effect of the {\it intentional} attack by deletion of the single most
connecting node or edge in the network
is reflected in the reduced size $S_{LC}$ of the largest connected component
after such an attack. The probability $P_{break}$
for a network to break up following the removal of a randomly selected
node or edge shows how sensitive it is to a random,
non-intentional failure.
}
\label{fig:phasePlane}
\end{figure}

Figure~\ref{fig:phasePlane}a shows the possible values of  $\cal{F}$ for
hierarchies and anti-hierarchies for $\gamma \in (2,3)$.
One can see that even the networks of narrow degree distribution can be organized
hierarchically (see upper limit for $\cal F$ for $\gamma=3$)
as well as networks of broad degree could be rearranged
to suppress ''self-hierarchical'' features
(see lower limit for  $\cal F$ for $\gamma=2$).

In Figure~\ref{fig:phasePlane}a we summarize
the results of our study of real and random scale-free
networks by displaying the hierarchical fraction ${\cal F}$ observed in
real world networks (black dots) relative to its value
for the random scale-free networks with the corresponding value of
$\gamma$ (solid line). As discussed above the
Internet, e-mail and CEO networks are about as hierarchical as
their random scale-free counterparts, while that of
protein-protein interactions in yeast is significantly
anti-hierarchical. Dark shaded regions in this figure correspond
to the range between maximally hierarchical and respectively
anti-hierarchical networks for a given value of $\gamma$.

Another interesting aspect is the impact the hierarchical structure
has on the overall robustness of the network.
In Figure 4b
we show the average size of the largest connected component $S_{LC}$
of the scale-free network with $\gamma=2.5$ and $N=1000$ after the
intentional attack, consisting of choosing and removing a single edge (left column)
or node (right column) in such a way as to minimize the $S_{LC}$ in the
resulting network. The smaller is the average $S_{LC}$ the more
vulnerable is the network with respect to attacks.
To characterize the robustness of a network with respect to
random failures we specify the likelihood $P_{break}$ that a removal of the single
node/edge disconnects the network.
We find that anti-hierarchical topologies are most vulnerable
with respect to attacks on their edges while
hierarchical topologies are sensitive to node attacks.
Apart from that, hierarchies are most vulnerable to
random failures.

In summary, we have discussed hierarchical organization
manifested in topology of complex networks, and demonstrated how
it can be used to characterize possible network architectures beyond
the degree distribution of their nodes.
We quantified the
hierarchal structure as the fraction of shortest paths that are
also hierarchical. It was found that this quantity approaches its maximum value
for marginally divergent scale-free networks $\gamma \leq 2$.
It was also shown that anti-hierarchy is naturally related to modular
features of networks. Finally we found that
hierarchal as well as anti-hierarchical network topologies
have implications for signaling and robustness against
various types of attacks and malfunctions, with anti-hierarchies
being quite reliable against
the most types of perturbations.

\indent
{\bf Acknowledgments:} We thank Aspen Center for Physics
for hospitality.
Work at Brookhaven National Laboratory was carried
out under Contract No. DE-AC02-98CH10886, Division
of Material Science, U.S. Department of Energy.
\vskip -0.5cm

\end{document}